\documentclass[aps,prl,showpacs,twocolumn,amsmath,10pt,nofootinbib,superscriptaddress,floatfix]{revtex4-1}

\usepackage{hyperref}
\usepackage{graphicx}
\usepackage{times}
\usepackage{color}
\usepackage{bm} 
\usepackage{braket} 
\usepackage{mwe}

\newcommand{\mN}{m_N}
\newcommand{\md}{m_d}
\newcommand{\mpi}{M_\pi}

\usepackage{enumitem}

\begin{document}
\title{Extraction of the neutron charge radius from a precision\\  
calculation of the deuteron structure radius}

\author{A.~A.~Filin}
\affiliation{Ruhr-Universit\"at Bochum, Fakult\"at f\"ur Physik und Astronomie,
Institut f\"ur Theoretische Physik II,
  D-44780 Bochum, Germany}
\author{V.~Baru}
\affiliation{Helmholtz-Institut f\"ur Strahlen- und Kernphysik and Bethe Center for Theoretical Physics, Universit\"at Bonn, D-53115 Bonn, Germany}
\affiliation{Institute for Theoretical and Experimental Physics NRC “Kurchatov Institute”, Moscow 117218, Russia}
\affiliation{P.N. Lebedev Physical Institute of the Russian Academy of Sciences, 119991, Leninskiy Prospect 53, Moscow, Russia}
\author{E.~Epelbaum}
\affiliation{Ruhr-Universit\"at Bochum, Fakult\"at f\"ur Physik und Astronomie,
Institut f\"ur Theoretische Physik II,
  D-44780 Bochum, Germany}
\author{H.~Krebs}
\affiliation{Ruhr-Universit\"at Bochum, Fakult\"at f\"ur Physik und Astronomie,
Institut f\"ur Theoretische Physik II,
D-44780 Bochum, Germany}
\author{D.~M\"oller}
\affiliation{Ruhr-Universit\"at Bochum, Fakult\"at f\"ur Physik und Astronomie,
Institut f\"ur Theoretische Physik II,
  D-44780 Bochum, Germany}
\author{P.~Reinert}
\affiliation{Ruhr-Universit\"at Bochum, Fakult\"at f\"ur Physik und Astronomie,
Institut f\"ur Theoretische Physik II,
  D-44780 Bochum, Germany}

\begin{abstract}
We present a high-accuracy
calculation of the deuteron structure radius in chiral effective
field theory. Our analysis employs the state-of-the-art semilocal
two-nucleon potentials and takes into account two-body contributions
to the charge density operators up to fifth order in the chiral
expansion.  The   strength of the fifth-order short-range two-body
contribution to the charge density operator is adjusted to
the experimental data on the deuteron charge form factor.
A detailed error analysis is performed by propagating the
statistical uncertainties of the low-energy constants entering the
two-nucleon potentials and by estimating errors from  the  truncation
of the chiral
expansion as well as from uncertainties in the nucleon form factors.
Using the predicted value for the deuteron structure radius together
with  the very accurate atomic data for the difference of the
deuteron and proton charge radii we, for the first time, extract
the charge radius of the neutron from light nuclei.
The extracted value reads $r_n^2 = - 0.106 \substack{ +0.007\\ -0.005\\} \, \text{fm}^2$ and
its magnitude
is about $1.7\sigma$ smaller than the current value given by the Particle Data Group.
In addition, given the high accuracy of the calculated  deuteron charge form factor and
its careful and systematic error analysis, our results open the way
for an accurate
determination of the nucleon form factors from elastic electron-deuteron
scattering data measured at the Mainz Microtron and other experimental facilities.
\end{abstract}


\maketitle

The tremendous progress in atomic spectroscopy achieved  
in the last decade led to a series  of  high-precision measurements of
the energy-level shifts in light atomic systems
which are  important for understanding   the structure of light nuclei  and their  charge distributions.
 In particular, a series of  extremely precise measurements of the
hydrogen-deuterium 1S-2S isotope shift (see Ref.~\cite{Jentschura:2011NOTinHep} for the latest update)
accompanied with an accurate theoretical QED analysis up through $O(\alpha^2)$
resulted in the  extraction of the deuteron-proton mean-square
charge radii difference~\cite{Pachucki:2018yxe} 
\begin{equation}
  r_d^2 - r_{p}^2 = 3.820 70(31) \text{fm}^2.
  \label{Eq:rd-rp}
\end{equation}
Because of its very high accuracy, this difference provides a tight link between $r_d$ and $r_p$ and thus
is important  in connection with the light nuclear charge radius
puzzle.
For many years,  the values for $r_p$ extracted from  electron and
muon experiments showed more than a 5$\sigma$ discrepancy~\cite{Pohl:2013yb}.
The very recent atomic hydrogen measurements~\cite{Beyer:2017gug,Bezginov:2019mdi},
however, claim consistency with the analogous muonic hydrogen experiments.
The  recommended value for the proton root-mean-square charge
radius has been changed to $r_p=0.8414(19)$~fm in the latest  CODATA-2018 update~\cite{CODATA2018}, and the deuteron
charge radius was updated accordingly,  by virtue of the  difference
in Eq.~\eqref{Eq:rd-rp}.
The updated CODATA deuteron charge radius  is only  1.9$\sigma$ larger
than the spectroscopic measurement on the muonic deuterium~\cite{Pohl1:2016xoo}
but still 2.9$\sigma$ smaller than the $r_d$
value from electronic deuterium spectroscopy~\cite{Pohl:2016glp}.
 
From the nuclear physics perspective, the charge radius of the
deuteron provides access to the deuteron internal structure   through
its structure radius, which is obtained from $r_d^2$ by subtracting
the contributions from the individual nucleons and the relativistic
(Darwin-Foldy)  correction,
\begin{eqnarray}
  r_{\rm str}^2 = r_d^2 - r_{p}^2 - r_{n}^2  - \frac{3}{4 m_p^2},
  \label{Eq:r_str}
\end{eqnarray}
where $m_p$ is the proton mass and $r_{n}^2$ is the neutron mean-square charge
radius. Traditionally, this relation is used to determine
$r_{\rm str}^2$ assuming that $r_d^2 - r_{p}^2$ and $r_{n}^2$ are
known.
The current value for the neutron  charge radius quoted by the PDG is
based on  measurements of the neutron-electron scattering length
in four different experiments carried out in 1973--1997
on   $^{208}$Pb,  $^{209}$Bi  and other heavy targets.  The world
average gives   $r_{n}^2 = -0.1161(22) \text{fm}^2$,  where the estimated error
was increased by a scaling factor of $1.3$~\cite{Tanabashi:2018oca}.
Nevertheless, the spread in the results on Pb and Bi  is significantly larger than even the  
increased uncertainty quoted by the PDG, which suggests  
that the error for the neutron mean-square charge radius might be
underestimated~\cite{Mitsyna:2009zz}.

With the recent advances in chiral effective field theory ($\chi$EFT),
theoretical analyses of low-energy few-nucleon reactions and nuclear
structure enter the precision
era~\cite{Epelbaum:2008ga,Machleidt:2011zz,Epelbaum:2019jbv}.
In this Letter we demonstrate that by employing  the nuclear forces
and currents derived up through fifth order in $\chi$EFT,  a very accurate determination of $r_{\rm str}$
is becoming possible from the analysis of  the deuteron
charge form factor (FF).  Equipped with this result
and using the information from the hydrogen-deuterium
isotope shift measurements given in
Eq.~\eqref{Eq:rd-rp}, we use the relation~\eqref{Eq:r_str} to extract,
for the first time, the neutron mean-square charge radius from the lightest atoms.

The electromagnetic FFs of the deuteron certainly belong to the most
extensively studied observables in nuclear physics, see
Refs.~\cite{Garcon:2001sz,Gilman:2001yh,Marcucci:2015rca} for review articles.
A large variety of theoretical approaches ranging from nonrelativistic quantum mechanics to covariant models have been
applied to this problem since the 1960s; see
Ref.~\cite{Phillips:2003pa} for an overview. The electromagnetic
structure of the deuteron has also been investigated in the framework of
pionless EFT~\cite{Chen:1999tn} and
$\chi$EFT~\cite{Phillips:1999am,Walzl:2001vb,Phillips:2003jz,Phillips:2006im,Valderrama:2007ja,Piarulli:2012bn,Epelbaum:2013naa}.
It is therefore crucial to emphasize the essential new aspects of the
current investigation.
\begin{enumerate}[itemsep=1pt,topsep=2pt,leftmargin=0.4cm]
\item[--]
    For the first time the calculation of the  deuteron charge FF is pushed to fifth order (N$^4$LO) in $\chi$EFT.
    This is achieved by
    (i) using the currently most accurate and precise
      $\chi$EFT two-nucleon (2N) potentials from Ref.~\cite{Reinert:2017usi}  and
    (ii) taking into account the short-ranged contribution to the two-body
      charge density operator at N$^4$LO.
\item[--]
    The two-body charge density is regularized consistently with the
    2N potential using the improved approach of Ref.~\cite{Reinert:2017usi}, which
    maintains the long-range interactions.
    The residual cutoff dependence of our results
    is verified to be well within
    the truncation uncertainty.
  \item[--]
    We employ the most up-to-date parametrizations of the nucleon FFs
    from the global analysis of  experimental data~\cite{Ye:2017gyb,Ye:smallrp}.
    To estimate the corresponding systematic uncertainty,
    we also  use the results from
    the dispersive analyses of
    Refs.~\cite{Belushkin:2006qa,Lorenz:2012tm,Lorenz:2014yda}, which
    incorporate constraints from unitarity and analyticity and predict the small proton radius consistent with the CODATA-2018 recommended value~\cite{CODATA2018}.   
\item[--]
    A thorough analysis of various types of uncertainty in
    the calculated deuteron FFs and the structure radius is performed.
\end{enumerate}

{\it Framework.}
In the Breit frame, the deuteron charge form factor is expressed in
terms of the matrix elements of the electromagnetic current,  convolved
with the deuteron wave functions as
\begin{eqnarray} \label{eq:defgc}
G_{\rm C}(Q^2) = \frac{1}{3e}  \frac{1}{2P_0}  
\sum_{\lambda} \braket{P',\lambda|J_B^0|P,\lambda},  \\
 \frac{1}{2P_0}\Braket{P', \lambda' | J_B^\mu| P, \lambda }  =
 \!\!\int \frac{d^3 l_1}{{(2\pi)}^3}  \frac{d^3 l_2}{{(2\pi)}^3} \times \\
 \nonumber
  \, \psi^\dagger_{\lambda'}\left(\bm{l_2} +\frac{\bm{k}}{4}, \bm{v_B}\right )
  \,  
   J^\mu_B \,
  \psi_{\lambda}\left(\bm{l_1} -\frac{\bm{k}}{4}, -\bm{v_B}\right ),
\label{Eq:ME}
\end{eqnarray}
where $e$ is the magnitude of electron charge, $J^\mu_B$ is the four-vector current calculated in the Breit
frame, $\psi_\lambda$ is the deuteron wave function with polarization $\lambda$ and the  deuteron in the
final (initial) state moves with the
velocity $\bm{v}_B $ ($-\bm{v}_B $) with  $\bm{v}_B =   {\bm
  k}/(2\sqrt{\bm{k}^2/4+\md^2}) =   \hat{\bm{k}} \sqrt{\eta/(1+\eta)}$
along the photon momentum.
The relativistic corrections to the deuteron wave functions related to
the motion of the initial and final deuterons are included along the
line of Ref.~\cite{Schiavilla:2002fq}.
Furthermore,    denoting the photon momentum $k = (0, \bm{k})$  (with
$Q^2=-k^2\ge0$) and the deuteron mass $m_d$, the deuteron initial and
final momenta  read $P = \left( P_0, -\bm{k}/2 \right)$ and $ P' =
\left( P_0, +\bm{k}/2 \right) $, respectively,
with $  P_0 =   \md \sqrt{1 + \eta}$  and $\eta = Q^2/{(2m_d)}^2$.
The deuteron charge radius is defined as  follows:
\begin{eqnarray}
  r_d^2 = (-6) {\left.\frac{\partial G_{\rm C} (Q^2)}{\partial Q^2} \right|}_{Q^2=0}.
\end{eqnarray}

The calculation of the deuteron FFs requires two important ingredients
which need to be derived in a consistent manner, namely the nuclear wave
functions and the electromagnetic currents.
The employed deuteron wave functions  are
calculated from the state-of-the-art  $\chi$EFT 2N potentials
of Ref.~\cite{Reinert:2017usi} which are among the most precise
interactions on the market.
Among many appealing features of these
interactions, we especially benefit from
a simple regularization scheme for the pion exchange contributions
which (i) maintains the
long-range part of the interaction, (ii) is applied in  momentum space
and (iii) allows for a straightforward
generalization to
current operators and many-body forces
at tree level.

The nuclear electromagnetic charge and current operators have been
recently worked out to N$^3$LO in $\chi$EFT
using the method of unitary transformation~\cite{Kolling:2009iq,Kolling:2011mt,Krebs:2019aka}
by our group and employing time-ordered perturbation theory~\cite{Pastore:2008ui,Pastore:2009is,Pastore:2011ip} by the JLab-Pisa group, see also
Ref.~\cite{Park:1995pn} for an early study along this
line.  The derivation of the electromagnetic currents and nuclear forces is carried out using
the Weinberg power counting   based on  the expansion parameter $Q =
p/ \Lambda_b$ with  $p \sim M_\pi$ being a characteristic soft momentum scale (with $M_{\pi}$ denoting the   pion mass) and  $\Lambda_b$
referring to the breakdown scale of the chiral expansion.
This implies that the  contributions to the charge  operators relevant for our study
appear at orders $Q^{-3}$ (LO),  $Q^{-1}$ (NLO),  $Q^{0}$ (N$^2$LO),
$Q^{1}$ (N$^3$LO) and $Q^{2}$ (N$^4$LO). To the  order we are working, the single-nucleon
contribution to the charge density operator in the kinematics $N(p) + \gamma(k) \to N(p^\prime)$
takes a well-known form (see, e.g., Ref.~\cite{Krebs:2019aka} and
references therein)
\begin{equation}
  \label{eq:rho1n}
  \rho_\text{1N} = e \left( 1 - \frac{{\bm{k}}^2}{8 \mN^2} \right) G_{\rm E} ({\bm{k}}^2)
  + i e \frac{G_{\rm ME}}{4 \mN^2} (\bm{\sigma} \cdot \bm{k} \times \bm{k_1}),
\end{equation}
where $\bm{k_1} = (\bm{p} +  \bm{p}')/2$,   
$G_{\rm E}({\bm{k}}^2)$ and $G_{\rm M}({\bm{k}}^2)$ are the electric and
magnetic form factors of the nucleon, $G_{\rm ME} :=2 G_{\rm M} ({\bm{k}}^2) - G_{\rm E} ({\bm{k}}^2)$, and $m_N$ denotes the nucleon mass.
The term $e G_{\rm E}$ on the rhs of Eq.~\eqref{eq:rho1n}  emerges at LO, while all other terms start to contribute at N$^3$LO.
Contributions to the two-body charge density
first appear at N$^3$LO from one- and two-pion exchange
diagrams, see Ref.~\cite{Kolling:2011mt} for explicit expressions.
Most of them are of the isovector type and,
therefore, do not contribute to the deuteron FFs. The only N$^3$LO
operator relevant for our study, to be denoted as
$\rho_\text{2N}^{1\pi} $, is a relativistic correction to the one-pion
exchange. It is  proportional to unobservable
phases $\bar{\beta}_8$ and $\bar{\beta}_9$ which parametrize
the unitary ambiguity of the long-range nuclear forces and currents at N$^3$LO.
In contrast to nuclear potentials,
observable quantities such as e.g.~the form factors must, of course,
be independent of the choice of $\bar{\beta}_8$, $\bar{\beta}_9$ and
other off-shell parameters. This can be achieved only by using
\emph{off-shell consistent} expressions for the nuclear forces and
currents.
Specifically, to preserve consistency with semilocal 2N potentials of
Ref.~\cite{Reinert:2017usi},
the one-pion exchange charge density has to be evaluated using the so-called minimal nonlocality choice with
$\bar{\beta}_8=1/4$ and  $  \bar{\beta}_9=-1/4$.
Although the pionic  
contributions to the isoscalar charge density at N$^4$LO have not been
worked out yet, the complete expression for the contact
operators at  N$^4$LO reads~\cite{Phillips:2016mov,our_long_paper}  (the contact term relevant for the quadrupole moment of the deuteron was first derived  in Ref.~\cite{Chen:1999tn}) 
\begin{eqnarray}
  \label{eq:contactchargedensity}
  \rho_\text{2N}^\text{cont}\nonumber
  =2 e G_{\rm E}^{\rm S}({\bm{k}}^2) \left(
    A \, \bm{k}^2
  + B \, \bm{k}^2 (\bm{\sigma}_1 \cdot \bm{\sigma}_2)
  + C \, \bm{k} \cdot \bm{\sigma}_1  \bm{k} \cdot \bm{\sigma}_2
   \right)\!,
\end{eqnarray}
where three low-energy constants (LECs) $A, B$ and $C$ contribute to the deuteron charge FF in \emph{one} linear combination only, 
see Supplemental Material \cite{supplem} for details.

The isoscalar electric nucleon form factor, $G_{\rm E}^{\rm S}({\bm{k}}^2)$,
is included  in  the two-body operators to  account for a non-pointlike character of the NN$\gamma$ vertex.
The chiral expansion of the electromagnetic FFs of the nucleon is well known to
converge slowly as they turn out to be dominated by the contributions
of vector mesons~\cite{Kubis:2000zd,Schindler:2005ke}, which are not included as explicit degrees of
freedom in $\chi$EFT.  Therefore, to minimize the impact of the slow convergence
of the chiral expansion of the nucleon FFs on 2N observables,  we
employ  up-to-date
parametrizations of the nucleon FFs  from Ref.~\cite{Ye:smallrp}
as well as from  several dispersive analyses of Refs.~\cite{Belushkin:2006qa,Lorenz:2012tm,Lorenz:2014yda}.

The 2N charge density operators $\rho^{1\pi}_\text{2N}$ and $
\rho^\text{cont}_\text{2N}$ have to be derived using the same regulator as employed in the 2N potentials.
The regularization of the operators with the single pion propagator is worked out in Ref.~\cite{Reinert:2017usi} and
can be effectively written as a substitution:
\begin{eqnarray}
  \frac{1}{\bm{p}^2+ \mpi^2}
  \to \frac{1}{\bm{p}^2+ \mpi^2} \exp \left( - \frac{\bm{p}^2+ \mpi^2}{\Lambda^2} \right),
  \label{eq:PionPropRegPresc}
\end{eqnarray}
where $\Lambda$ is a fixed cutoff chosen consistently with the employed 2N potential  in the range %
of $400$--$550$~MeV.  The prescription for regularizing the squared pion propagator
consistent with the approach used in~\cite{Reinert:2017usi}
can be obtained from Eq.~\eqref{eq:PionPropRegPresc} by taking a
derivative with respect to $\mpi^2$.
To maintain consistency between $\rho^\text{cont}_\text{2N}$ and the
corresponding short-range terms in the 2N potential after
regularization,
we exploit the fact that both
can be generated from the same unitary transformation acting on
the single-nucleon charge density and the kinetic energy
term, respectively~\cite{our_long_paper}.

\begin{figure}
\begin{center}
\hspace*{-0.3cm}\includegraphics[width=8.7cm]{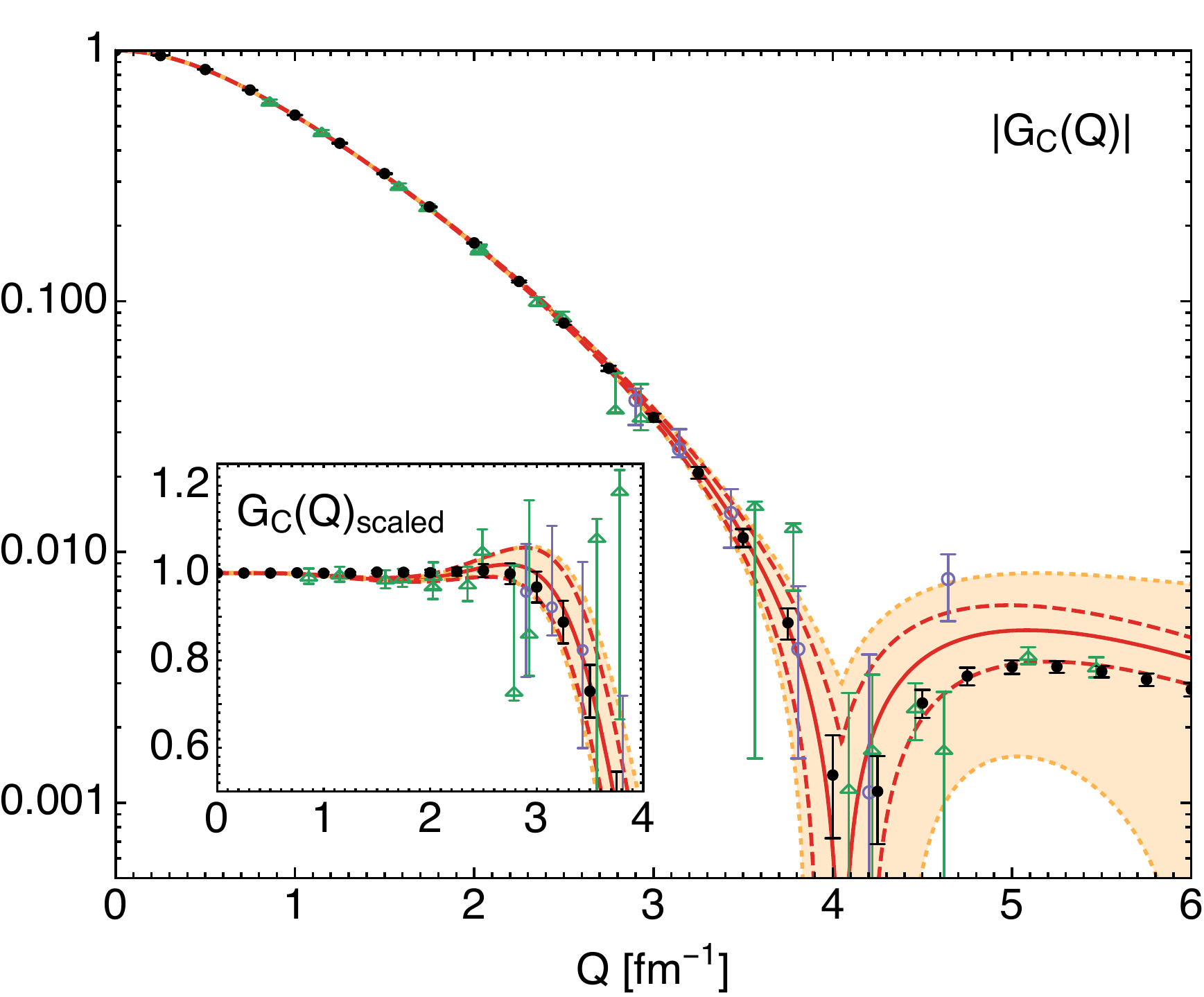}
\caption{\label{fig:GC_FFFit}
  (Color online)   Deuteron charge FF from the best fit  to data up to $Q=4$ fm$^{-1}$ evaluated for the cutoff  $\Lambda = 500$ MeV (solid red lines).
   Band between  dashed (red) lines corresponds to a $1\sigma$ error in the determination of
   the short-range contribution to the charge density operator at N$^4$LO.
  Light-shaded (orange dotted) band corresponds to
  the estimated  error ($68\%$  degree-of-belief) from truncation of the chiral expansion at N$^4$LO.
  Open violet circles and green triangles are experimental data from Ref.~\cite{Nikolenko:2003zq} and
Refs.~\cite{Abbott:2000ak,Abbott:2000fg}, respectively.
  Black solid circles correspond to the parametrization of the deuteron FFs from Refs.~\cite{Marcucci:2015rca,Sick:priv}
  which  is not used in the fit and shown just for comparison. The  rescaled charge FF of the deuteron, $G_\text{C}{(Q)}_{\rm scaled}$, as defined in  Ref.~\cite{Marcucci:2015rca}, is shown on a linear scale.
}
\end{center}
\end{figure}

{\it Results and discussions.}  The calculated deuteron FF at N$^4$LO,
$G^{\rm th}_{\rm C} (Q)$, involves one unknown parameter (a
combination of the LECs from $\rho^\text{cont}_\text{2N}$), which is
extracted from a fit to the
world data for the  deuteron charge form factor $G^\text{exp}_\text{C}
(Q)$ from Refs.~\cite{Abbott:2000ak,Abbott:2000fg,Nikolenko:2003zq}.
Here and in what follows, the N$^4$LO results are obtained using
the N$^4$LO$^+$ 2N potentials, which include 4 
  sixth-order contact interactions in F waves and result in a nearly
  perfect description of 2N data from the Granada 2013 database~\cite{Perez:2013jpa} below the pion production threshold.

\begingroup
\squeezetable
\begin{table}[t]
\caption{\label{Tab:rstr_uncert} Deuteron structure radius squared predicted at
N$^4$LO in $\chi$EFT (1st column) and the individual contributions
to its uncertainty: from
the truncation of the chiral expansion (2nd),  the statistical error in the
short-range charge density operator extracted from $G_{\rm C} (Q^2)$
(3rd),  the errors from the
statistical uncertainty in $\pi$N LECs from the Roy-Steiner analysis of Ref.~\cite{Hoferichter:2015tha,Hoferichter:2015hva}
propagated  through the variation in the deuteron wave functions (4th),   the errors from the
statistical uncertainty in 2N LECs extracted from the Granada
  2013 2N database  in Ref.~\cite{Reinert:2017usi} (5th), the error from  the
choice of the maximal energy   in the fit (6th)
as well as the total uncertainty evaluated using the sum of these numbers in quadrature (7th).
All numbers are given in fm$^2$.
}
\small
\begin{ruledtabular} 
\begin{tabular*}{0.48\textwidth}{@{\extracolsep{\fill}}ccccccc}
{$r_{\rm str}^2$ } & {{\rm Truncation}} &  $\rho^{\rm cont}_\text{2N}$ &
                                                                    {$\pi$N LECs} & {2N LECs}                                   & $Q$ range &   {Total} \\
\hline
3.8933                                 & $\pm$0.0032         & $\pm$0.0037 & $\pm$0.0004             &     $ \substack{+0.0010\\ -0.0047}$       & $\pm$0.0017                                    & $ \substack{+0.0053\\ -0.0070}$               \\
\end{tabular*}
\end{ruledtabular}
\end{table}
\endgroup
The function $\chi^2$  to be minimized in the fit   is defined as follows
\begin{eqnarray}\label{chisq}
  \chi^2 = \sum_i \frac{{(G^{\rm th}_{\rm C} (Q_i) - G^\text{exp}_{\rm C} (Q_i))}^2}{\Delta G_{\rm C} {(Q_i)}^2},
\end{eqnarray}
where following Refs.~\cite{Carlsson:2015vda,Wesolowski:2018lzj}
the uncertainty $\Delta G_{\rm C} (Q_i)$ besides the experimental errors also takes into
account theoretical uncertainties from the truncation of the chiral
expansion estimated using the Bayesian approach
and from the nucleon form factors, as given in Refs.~\cite{Ye:2017gyb,Ye:smallrp},
added in quadrature. Throughout this analysis, we employ the
Bayesian model $\bar C_{0.5-10}^{650}$ specified in
Ref.~\cite{Epelbaum:2019zqc} and assume the characteristic momentum
scale to be given by $|\bm{k}|/2$~\cite{Phillips:2006im}.
The results for the  deuteron charge FF from the best fit  to data up
to $Q=4$ fm$^{-1}$, evaluated for the cutoff  $\Lambda = 500$ MeV, are
visualized in Fig.~\ref{fig:GC_FFFit}
together with the N$^4$LO truncation errors and statistical uncertainty
of the strength of $\rho^{\rm cont}_\text{2N}$.
We have verified that the  cutoff variation in the range of $\Lambda = 400
\ldots 550$~MeV  yields results lying well within the
truncation error band and that the fits of a similar quality can be
obtained by employing the nucleon
FFs from the dispersive analyses of
Refs.~\cite{Belushkin:2006qa,Lorenz:2012tm,Lorenz:2014yda}
, see
Ref.~\cite{our_long_paper} for a detailed discussion of various
uncertainties.  
The fit to data allows us   to accurately extract the unknown linear
combination of LECs
entering the charge density operator at N$^4$LO and  thus  to make
a parameter-free prediction for  the structure radius of the deuteron,
which reads 
\begin{eqnarray}\label{rsrt_th}
  r_{\rm str} = 1.9731 \substack{+0.0013\\ -0.0018}\ \text{fm},
\end{eqnarray}
with the individual contributions to the uncertainty given in
Table~\ref{Tab:rstr_uncert}.  To make this
uncertainty estimate   conservatively,  
the truncation error is actually included twice: (i) by performing the
Bayesian analysis for  $r_{\rm str}^2$ following the approach of
Ref.~\cite{Epelbaum:2019zqc} and (ii) through the statistical
uncertainty in the short-range charge density extracted from the fit
to $G_{\rm C}^\text{exp} (Q^2)$
using  Eq.~\eqref{chisq}. Furthermore, we developed a phase-equivalent
version of the 2N potential using a different choice of the
unobservable phases $\bar{\beta}_8=  \bar{\beta}_9=1/2$ leading
to $\rho^{1\pi}_\text{2N} = 0$. Repeating the analysis for this
choice of $\bar {\beta}_{8,9}$, the value of $r_{\rm str}$ is found to
agree with the one in Eq.~(\ref{rsrt_th}) to all given figures.
The  structure radius  
is also robust with respect to data
used in the fit:  had we used the parametrization of data by Sick~\cite{Marcucci:2015rca,Sick:priv}
instead of experimental data, we would have arrived at essentially
the same result.  For the sake of completeness, we also present the results of the
order-by-order calculations for  $r_{\rm str}$ (in units of fm) including the truncation error from the Bayesian analysis,
$1.9  \pm 0.4$ (LO),
$1.97 \pm 0.03$ (NLO),
$1.969 \pm 0.007$ (N$^2$LO),
$1.969 \pm 0.002$ (N$^3$LO),
$1.9731 \pm 0.0008$ (N$^4$LO).
 It is important to keep in mind that these numbers are obtained
 without relying on the chiral expansion of the nucleon form
 factors.

Relying on
our theoretical prediction for the structure radius, we  are now in the
position to predict the neutron charge radius from
Eqs.~\eqref{Eq:rd-rp}, \eqref{Eq:r_str} and~\eqref{rsrt_th}, which gives
\begin{equation}
 \label{Eq:neut_rad}
 r_n^2 = - 0.106 \substack{ +0.007\\ -0.005\\} \, \text{fm}^2.
\end{equation}
This value is   $1.7\sigma$ smaller than the one given by the PDG~\cite{Tanabashi:2018oca}.

In summary,  we presented   a comprehensive analysis of the deuteron
charge form factor up to fifth order in $\chi$EFT.
The only unknown parameter enters
the  short-range 2N contribution  to the charge density operator and is
determined from the best fit to the deuteron charge form factor.
Equipped with this information,  we make a parameter-free
prediction for the structure radius of the deuteron and perform a
thorough analysis of various kinds of uncertainty.
The high-accuracy calculation of the structure radius, together with
the high-precision measurement of the hydrogen-deuterium 1S-2S
isotope shift~\cite{Jentschura:2011NOTinHep}, have allowed us
to extract the neutron charge radius.

Although it is natural to expect that the two-pion exchange
contributions to the charge density at N$^4$LO, which have not yet
been worked  out,
are largely saturated by the short-range contributions included in
this analysis,
the complete $\chi$EFT calculation at this order would
allow for an  additional test of the estimated theoretical uncertainty.

The results for the deuteron charge FF presented here
pave the way for  an accurate determination of the isoscalar nucleon FF
by (re)analyzing the experimental data
on elastic electron-deuteron scattering  at MAMI (see e.g.~Ref.~\cite{Schlimme:2016wmj} for the new measurement
of the elastic $ed$ scattering cross section at 0.24 fm$^{-1}\!\le \!
Q \le$ 2.7 fm$^{-1}$ at MAMI), Saclay~\cite{Platchkov:1989ch} and other facilities.

{\bf Acknowledgments.}  
 We are  grateful to U.-G.~Mei\ss ner for a careful reading of the manuscript and valuable comments and to Z.~Ye for providing us with the unpublished results for the nucleon form factors from Ref.~\cite{Ye:smallrp}.
 We also thank H.-W.~Hammer for providing us with the parametrization of the nucleon form factors from Ref.~\cite{Belushkin:2006qa} and   I.~Sick for  the parametrization of the deuteron form factors
from Ref.~\cite{Marcucci:2015rca}.  We are grateful to M.  Hoferichter and J.~Ruiz de Elvira   for the information on the central values and covariance matrix of the N$^4$LO $\pi$N LECs from the Roy-Steiner analysis.
This work was supported in part by DFG and NSFC through funds provided to the Sino-German CRC 110 ``Symmetries and
 the Emergence of Structure in QCD'' (NSFC Grant No. 11621131001, Grant No.~TRR110), the BMBF (Grant No. 05P18PCFP1)  and the Russian Science Foundation (Grant No.~18-12-00226).

\end{document}


\title{Supplemental material for `Extraction of the neutron charge radius from a precision
calculation of the deuteron structure radius'}

\author{A.~A.~Filin}
\affiliation{Ruhr-Universit\"at Bochum, Fakult\"at f\"ur Physik und Astronomie,
Institut f\"ur Theoretische Physik II,
  D-44780 Bochum, Germany}
\author{V.~Baru}
\affiliation{Helmholtz-Institut f\"ur Strahlen- und Kernphysik and Bethe Center for Theoretical Physics, Universit\"at Bonn, D-53115 Bonn, Germany}
\affiliation{Institute for Theoretical and Experimental Physics, B. Cheremushkinskaya 25, 117218 Moscow, Russia}
\affiliation{P.N. Lebedev Physical Institute of the Russian Academy of Sciences, 119991, Leninskiy Prospect 53, Moscow, Russia}
\author{E.~Epelbaum}
\affiliation{Ruhr-Universit\"at Bochum, Fakult\"at f\"ur Physik und Astronomie,
Institut f\"ur Theoretische Physik II,
  D-44780 Bochum, Germany}
\author{H.~Krebs}
\affiliation{Ruhr-Universit\"at Bochum, Fakult\"at f\"ur Physik und Astronomie,
Institut f\"ur Theoretische Physik II,
D-44780 Bochum, Germany}
\author{D.~M\"oller}
\affiliation{Ruhr-Universit\"at Bochum, Fakult\"at f\"ur Physik und Astronomie,
Institut f\"ur Theoretische Physik II,
  D-44780 Bochum, Germany}
\author{P.~Reinert}
\affiliation{Ruhr-Universit\"at Bochum, Fakult\"at f\"ur Physik und Astronomie,
Institut f\"ur Theoretische Physik II,
  D-44780 Bochum, Germany}

\maketitle

\section{Contributions from the single-nucleon charge density operators}

For completeness, here we provide the expressions for the single-nucleon charge density operators which contribute to the deuteron charge form factor (FF) and charge radius.
Our kinematics for single-nucleon charge density is $N(\bm{p}) + \gamma(\bm{k}) \to N(\bm{p}^\prime)$.
The single-nucleon charge density operators (see e.g.~a review~\cite{Krebs:2019aka}) which are relevant for the calculation of the deuteron form factors  up to N$^4$LO can be grouped into three terms
\begin{eqnarray}
	\label{eq:rho_decomp}
 	\rho_\text{1N} = \rho^{\rm Main}_\text{1N} +  \rho_\text{1N}^\text{DF}+ \rho_\text{1N}^\text{SO},
\end{eqnarray}
where the terms ``Main'', Darwin-Foldy (DF) and spin-orbit (SO) read
\begin{eqnarray}
	  \label{eq:rho1Nmain}
  \rho_\text{1N}^\text{Main} = e G_\text{E}({\bm{k}}^2), \hspace{1.cm}
    \rho_\text{1N}^\text{DF} = e \left(  - \frac{{\bm{k}}^2}{8 \mN^2} \right) G_\text{E}({\bm{k}}^2), \hspace{1.cm}
    \rho_\text{1N}^\text{SO} =
   i e \frac{2 G_\text{M}({\bm{k}}^2) - G_\text{E}({\bm{k}}^2)}{8 \mN^2} \bm{\sigma} \cdot \bm{k} \times (\bm{p +p'}),
\end{eqnarray}
where $e$ is the electron charge, $m_N$ stands for the average nucleon mass, and the vector
$\bm{\sigma}$  denotes the  Pauli matrices in the spin space.
The nucleon electric and magnetic form factors are defined as
\begin{eqnarray}
  G_\text{E}({\bm{k}}^2) = G_\text{E}^p({\bm{k}}^2) \frac{1 + \tau_3}{2} + G_\text{E}^n({\bm{k}}^2) \frac{1 - \tau_3}{2},
  \qquad
  G_\text{M}({\bm{k}}^2) = G_\text{M}^p({\bm{k}}^2) \frac{1 + \tau_3}{2} + G_\text{M}^n({\bm{k}}^2) \frac{1 - \tau_3}{2},
\end{eqnarray}
where $G_\text{E(M)}^{p(n)}$ denote the individual proton and neutron electric and magnetic form factors, respectively,
and $\tau$ denotes the Pauli matrices in the isospin space.
The deuteron form factors depend only on the isoscalar combinations of the nucleon form factors, namely
\begin{eqnarray}
  G_\text{E}^\text{S}({\bm{k}}^2) := G_\text{E}^p({\bm{k}}^2) + G_\text{E}^n({\bm{k}}^2),
  \qquad
  G_\text{M}^\text{S}({\bm{k}}^2) := G_\text{M}^p({\bm{k}}^2) + G_\text{M}^n({\bm{k}}^2).
  \label{eq:isoscalarFFs}
\end{eqnarray}
The one-body contributions to the deuteron charge form factor  
 evaluated by calculating the expectation values of
$\rho^{\rm Main}_\text{1N}$, $\rho_\text{1N}^\text{DF}$, $\rho_\text{1N}^\text{SO}$
including a relativistic correction, $G_\text{C}^\text{Boost}({\bm{k}}^2)$, due to the motion of the initial and final deuterons (see e.g. Refs.~\cite{Schiavilla:2002fq,Phillips:2003pa} for details) read
\begin{eqnarray}
  G_\text{C}^\text{1N}({\bm{k}}^2) = G_\text{C}^\text{Main}({\bm{k}}^2)
    + G_\text{C}^\text{DF}({\bm{k}}^2)
    + G_\text{C}^\text{SO}({\bm{k}}^2)
    + G_\text{C}^\text{Boost}({\bm{k}}^2).
\end{eqnarray}

\section{Two-nucleon charge density: one-pion exchange operators at N$^3$LO}

In this section we provide the regularized expressions for the relevant (two-nucleon) one-pion exchange charge-density operators.
For two-nucleon charge-density operators we employ the following kinematics
\begin{eqnarray}\label{Eq:kinem}
  N(\bm{p}_1) + N(\bm{p}_2) + \gamma(\bm{k}) \to N(\bm{p}_1^\prime) + N(\bm{p}_2^\prime)
\end{eqnarray}
and define auxiliary three-momenta
$\bm{q}_1 := \bm{p}_1^{\,\prime} - \bm{p}_1$ and $\bm{q}_2 := \bm{p}_2^{\,\prime} - \bm{p}_2$.
In this notation, the
unregularized contribution to the isoscalar charge-density operator from one-pion exchange
reads~\cite{Kolling:2011mt}:
\begin{eqnarray}
  \label{eq:NNchargeDensityKolling}
  \rho_\text{2N}^{1\pi} &=&
	\frac{-3 e g_A^2}{16 F_\pi^2 \mN}
  \left(
  (1-2 \bar{\beta}_9)
  \frac{
    (\bm{\sigma}_1 \cdot \bm{k})
    (\bm{\sigma}_2 \cdot \bm{q}_2)
    }
    {\bm{q}_2^2 + \mpi^2}
  + (2 \bar{\beta}_8 - 1)
  \frac{
    (\bm{\sigma}_1 \cdot \bm{q}_2)
    (\bm{\sigma}_2 \cdot \bm{q}_2)
    (\bm{q}_2 \cdot \bm{k})
    }
    {{(\bm{q}_2^2 + \mpi^2)}^2}
    \right)
   + (1 \leftrightarrow 2),
\end{eqnarray}
where $\bar{\beta}_8$, $\bar{\beta}_9$ are unobservable unitary phases (see the main text and Ref.~\cite{Kolling:2011mt}),
$\gA$ is the axial-vector coupling constant of the nucleon,
$\fpi$ is the pion decay constant,
$\mpi$ is the pion mass
and $(1 \leftrightarrow 2)$ corresponds to the  term with interchanged nucleon lines.
Regularizing  Eq.~\eqref{eq:NNchargeDensityKolling} consistently with the 2N potential from Ref.~\cite{Reinert:2017usi}
gives
\begin{eqnarray}
  \rho_\text{2N}^{1\pi, \, \text{reg}} &=&
  (1-2 \bar{\beta}_9) G_\text{E}^\text{S}(\bm{k}^2)
  \frac{-3e g_A^2}{16 F_\pi^2 \mN}
  \frac{
    (\bm{\sigma}_1 \cdot \bm{k})
    (\bm{\sigma}_2 \cdot \bm{q}_2)
    }
    {\bm{q}_2^2 + \mpi^2}
    \exp \left( - \frac{\bm{q}_2^2+ \mpi^2}{\Lambda^2} \right)
  \nonumber
  \\
  && + (2 \bar{\beta}_8 - 1) G_\text{E}^\text{S}(\bm{k}^2)
  \frac{-3e g_A^2}{16 F_\pi^2 \mN}
    (\bm{\sigma}_1 \cdot \bm{q}_2)
    (\bm{\sigma}_2 \cdot \bm{q}_2)
    (\bm{q}_2 \cdot \bm{k})
  \nonumber
  \\
    && \times
    \left( \frac{1}{{(\bm{q}_2^2+ \mpi^2)}^2}  + \frac{1}{\Lambda^2 (\bm{q}_2^2+ \mpi^2)} \right)
    \exp \left( - \frac{\bm{q}_2^2+ \mpi^2}{\Lambda^2} \right)
   + (1 \leftrightarrow 2).
   \label{eq:NNchargeDensityKollingWithGESregularized}
\end{eqnarray}
where the value of the momentum-space cutoff $\Lambda$ should be taken consistently with that employed in  2N potential.
We emphasize that in order to obtain observables independent of the choice for $\bar{\beta}_8$ and $\bar{\beta}_9$,  it is important to include the nucleon form-factor $G_\text{E}^\text{S}(\bm{k}^2)$ in~\eqref{eq:NNchargeDensityKollingWithGESregularized}.

\section{Two-nucleon charge density: contact operators  at N$^4$LO}

The  isoscalar contact contribution to the 2N charge density operator at  N$^4$LO reads
\begin{eqnarray}
  \label{eq:contactchargedensity}
  \rho_\text{cont}
  =2 e G_\text{E}^\text{S}({\bm{k}}^2) \left(
    A \, \bm{k}^2
  + B \, \bm{k}^2 (\bm{\sigma}_1 \cdot \bm{\sigma}_2)
  + C \, \bm{k} \cdot \bm{\sigma}_1  \bm{k} \cdot \bm{\sigma}_2
   \right),
\end{eqnarray}
where $A$, $B$, and $C$ are low-energy constants (LECs).  It is important to realize that  the contribution of  
$ \rho_\text{Cont}$  to the deuteron charge FF involves only one linear combination of these LECs, namely 
$( A+B+\frac{C}{3} )$, see next section for the value extracted from the best fit to data.
After regularization which is  carried out consistently with the 2N potential, as explained in the main text, the contact operator  takes the form
\begin{eqnarray}
  \label{eq:contactchargedensityReg}
  \rho_\text{cont}^\text{reg}
  = 2 e G_\text{E}^\text{S}({\bm{k}}^2) \left(
    (A + B \,  (\bm{\sigma}_1 \cdot \bm{\sigma}_2))
    F_1 \left( \frac{\bm{p}_1-\bm{p}_2}{2}, \frac{\bm{p}'_1-\bm{p}'_2}{2}, \bm{k} \right)
  + C F_2 \left( \frac{\bm{p}_1-\bm{p}_2}{2}, \frac{\bm{p}'_1-\bm{p}'_2}{2}, \bm{k} \right)
   \right),
\end{eqnarray}
where the functions $F_1$ and $F_2$ are defined as
\begin{eqnarray}
  F_i (\bm{p}, \bm{p}', \bm{k}) &:=&
      E_i \left( \bm{p}-\frac{\bm{k}}{2}, \bm{p}' \right)
    + E_i \left( \bm{p}+\frac{\bm{k}}{2}, \bm{p}' \right)
    + E_i \left( \bm{p}'-\frac{\bm{k}}{2}, \bm{p} \right)
    + E_i \left( \bm{p}'+\frac{\bm{k}}{2}, \bm{p} \right),
\\
  E_1 \left( \bm{p}, \bm{p}' \right) &:=& \left( \bm{p}^2 - \bm{p}'^2 \right)
   e^{-\frac{\bm{p}^2 + \bm{p}'^2}{ \Lambda ^2}},
  \qquad
  E_2 \left( \bm{p}, \bm{p}' \right) := \left[
    (\bm{\sigma}_1 \cdot \bm{p}) (\bm{\sigma}_2 \cdot \bm{p}) -  (\bm{\sigma}_1 \cdot \bm{p}') (\bm{\sigma}_2 \cdot \bm{p}')
   \right]
   e^{-\frac{\bm{p}^2 + \bm{p}'^2}{ \Lambda ^2}}.
\end{eqnarray}
In analogy to the discussion in the previous section,
we also include $G_E^S({\bm{k}}^2)$ in Eqs.~\eqref{eq:contactchargedensity} and \eqref{eq:contactchargedensityReg}.

\section{Deuteron charge form factor and structure radius calculation}

The relevant contributions  to the deuteron charge form factor, deuteron charge radius and structure radius  read
\begin{eqnarray}
\label{eq:GCcontributions}
  G_\text{C}^\text{th}(Q^2) = G_\text{C}^\text{Main}(Q^2)
    + G_\text{C}^\text{DF}(Q^2)
    + G_\text{C}^\text{SO}(Q^2)
    + G_\text{C}^\text{Boost}(Q^2)
    + G_\text{C}^{1\pi}(Q^2)
    + G_\text{C}^\text{cont}(Q^2),
\end{eqnarray}
\begin{eqnarray}
  r_d^2 = r_{m}^2 + r_{p}^2 + r_{n}^2  + r_{\rm DF}^2  + r_{\rm SO}^2  + r_{1\pi}^2  + r_\text{cont}^2,
\end{eqnarray}
\begin{eqnarray}
  r_\text{str}^2 = r_{m}^2 + r_{\rm SO}^2  + r_{1\pi}^2  + r_\text{cont}^2,
\end{eqnarray}
where contributions with index ``$1 \pi$'' (``cont'') stem from the regularized 2N charge
density Eq.~\eqref{eq:NNchargeDensityKollingWithGESregularized} (Eq.~\eqref{eq:contactchargedensityReg}),
$r_m$ is the deuteron matter radius,
$r_p$ and $r_n$ denote the proton and neutron charge radii, respectively, and in the Breit frame $Q^2= \bm k^2$.
The relevant LEC,  $A+B+C/3$, from Eq.~\eqref{eq:contactchargedensityReg}  
 is extracted from the best fit to data on the deuteron charge FF  by minimizing the $\chi^2$ function defined as
\begin{eqnarray}
	\chi^2= \sum_i \frac{{\bigl(G^{\rm th}_\text{C}(Q^2_i) - G^\text{exp}_\text{C}(Q^2_i)\bigl)}^2}{\Delta G_\text{C}{(Q_i^2)}^2},
\end{eqnarray}
where following Refs.~\cite{Carlsson:2015vda,Wesolowski:2018lzj}
the uncertainty $\Delta G_{\rm C} (Q_i^2)$ besides the experimental errors also takes into
account theoretical uncertainties from the truncation of the chiral
expansion estimated using the Bayesian approach
and from the nucleon form factors, as given in Ref.~\cite{Ye:2017gyb,Ye:smallrp},
added in quadrature, namely
\begin{eqnarray}
	\Delta G_\text{C}{(Q_i^2)}^2 =\Delta G_\text{C}^\text{exp}{(Q_i^2)}^2 + \Delta G_\text{C}^\text{th,trunc}{(Q_i^2)}^2 +\Delta G_\text{C}^\text{th,nuclFF}{(Q_i^2)}^2.
\end{eqnarray}
The  value for the LEC  $A+B+C/3$   for the cutoff $\Lambda = 500$ MeV reads
\begin{eqnarray}
A+B+\frac{C}{3} = (-340 \pm 95) \text{ GeV}^{-5} =
(-0.8 \pm 0.2) \frac{1}{\fpi^{2} \Lambda_b^{3} },  
\end{eqnarray}
where the error corresponds to the 1$\sigma$ deviation  and $\Lambda_b=650 \text{ MeV}$
refers to the breakdown scale of the chiral expansion.

\section{Numerical values}
Here we specify the values of various constants and masses entering numerical calculations, namely,
$\hbar c = 1 = 0.197327$ GeV fm,
$\mN = 0.938918$ GeV,
$m_d = 1.87561$ GeV,
$\gA = 1.29$,
$\mpi = 0.13803$ GeV,
$\fpi = 0.0924$ GeV.
These values are  consistent with the ones employed in the 2N potential of Ref.~\cite{Reinert:2017usi}.